# Application of Joint Notch Filtering and Wavelet Transform for Enhanced Powerline Interference Removal in Atrial Fibrillation Electrograms

Miguel Martínez-Iniesta[1], Juan Ródenas[1], José J. Rieta[2], Raúl Alcaraz[1]

[1] Research Group in Electronic, Biomedical and Telecomm. Eng., Univ. of Castilla-La Mancha, Spain
[2] BioMIT.org, Electronic Engineering Department, Universitat Politecnica de Valencia, Spain

**Abstract**

*Analysis of intra-atrial electrograms (EGMs) nowadays constitutes the most common way to gain new insights about the mechanisms triggering and maintaining atrial fibrillation (AF). However, these recordings are highly contaminated by powerline interference (PLI) due to the large amount of electrical devices operating simultaneously in the electrophysiology laboratory. To remove this perturbation, conventional notch filtering has been widely used. However, this method adds artificial fractionation to the EGMs, thus concealing their accurate interpretation. Hence, the development of novel algorithms for PLI suppression in EGMs is still an unresolved challenge. Within this context, the present work introduces the joint application of common notch filtering and Wavelet denoising for enhanced PLI removal in AF EGMs. The algorithm was validated on a set of 100 unipolar EGM signals, which were synthesized with different noise levels. Original and denoised EGMs were compared in terms of a signed correlation index (SCI), computed both in time and frequency domains. Compared with the single use of notch filtering, improvements between 4 and 15% were reached with Wavelet denoising in both domains. As a consequence, the proposed algorithm was able to efficiently reduce high levels of PLI and simultaneously preserve the original morphology of AF EGMs.*

## 1. Introduction

Atrial fibrillation (AF) is one of the most relevant cardiovascular challenges in western countries [1], affecting approximately 1.5–2% of the general population [2]. Although it is the most common cardiac arrhythmia in clinical practice [2], its mechanisms are not fully known [3, 4]. This fact makes diagnosis and treatment of the arrhythmia poorly effective and complex, thus demanding more research efforts [5]. In fact, catheter ablation, which is currently considered the first-line therapy for AF, still does not provide clinically sufficient long-term success rates [6].

Hence, there is a need for more advanced processing and interpretation of cardiac electrophysiology systems [6], which use intra-atrial electrograms (EGMs) as the basis to determine the cardiac structures contributing to sustain the arrhythmia [7].

These EGMs are directly recorded from the heart, thus providing information about its electric status [8]. More precisely, these signals offer accurate indications about the time, direction and complexity of local atrial activations within the field of view of the recording electrodes [9]. Hence, characterization of these recordings is the best source of information to improve current knowledge on the mechanisms responsible for initiation and perpetuation of AF [10]. However, their acquisition is often disturbed by the presence of numerous sources of electric noise. Thus, in addition to the internal noise introduced by the recording systems as well as common baseline wandering from patient's respiration, other disturbances from muscular activity of the patient and the powerline interference (PLI) can also reduce the quality of the EGMs [11, 12].

This last perturbation is a typical environmental electromagnetic noise mainly generated by power cords radiation that can be induced to medical instrument signal cables and to the patient's body with an amplitude comparable to the physiological information [13]. Unfortunately, given that PLI is spectrally overlapped with the EGM, its removal is a challenging task. However, this aspect has still not received much attention and comercial recording systems only incorporate low-order simple notch filters, which remove important local cardiac components along with the PLI [12]. This work introduces the joint application of common notch filtering and Wavelet denoising to efficiently reduce high levels of PLI and simultaneously preserving the original morphology of AF EGMs.

## 2. Methods

### 2.1. Study database

With the aim of accurately quantifying the achieved PLI reduction as well as the signal morphology alteration, the proposed algorithm was validated on a set of 100 synthesized unipolar EGM recordings. They were generated with a length of 10 seconds making use of the methodology proposed by Oesterlin et al. [14]. To obtain recordings with signal-to-interference ratios (SIR) of 25, 20, 15, 10 and 5 dB, a sinusoidal interference with a main frequency of 50 Hz and its first two harmonics were added to the resulting EGMs. Note that random amplitude and frequency variations were considered in the generated PLI in order to mimic an interference as realistic as possible.

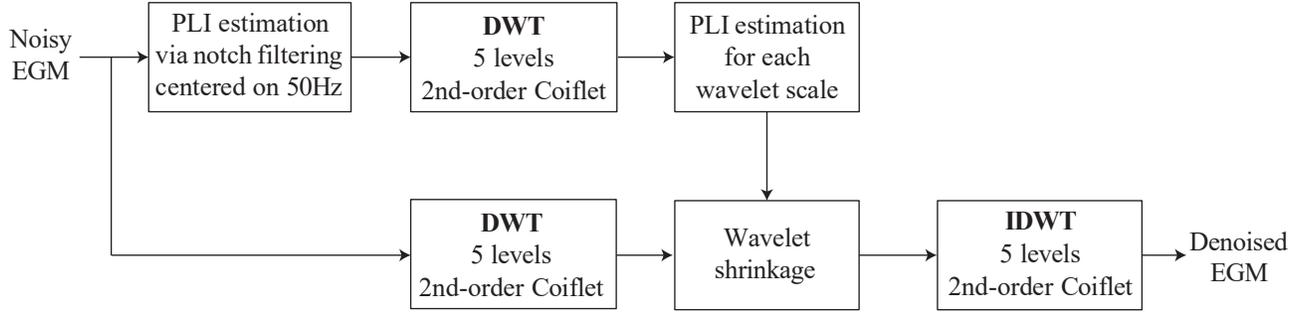

Figure 1. Block diagram summarizing the proposed DWT-based algorithm for PLI removal in EGM signals.

### 2.2. Wavelet denoising

Given its ability to successfully deal with non-stationary signals containing sharp spikes and discontinuous intervals, discrete Wavelet transform (DWT) has been widely used for the denoising of many physiological record- ings [15]. In brief, this tool allows to decompose a sig- nal at different time and frequency scales through a sim- ple implementation, which consists of a bank of low-pass and high-pass filters followed by decimation stages. Next, the resulting wavelet coefficients are thresholded to dis- cern physiological information from noise and the de- noised signal is reconstructed by computing the inverse DWT (IDWT).

In this approach the selection of a proper shrinkage threshold for each wavelet scale plays a key role for a successful denoising. Although well-established methods exist to obtain these cut-off points in the presence of white noise, they are not applicable to sinusoidal interference cancellation [16]. Hence, for PLI removal in EGM recordings, the perturbation amplitude was firstly estimated through a simple common notch filtering centered on the frequency of 50 Hz and with a bandwidth of 2 Hz, such as Figure 1 shows. Although the resulting sig- nal was still disturbed by EGM residua, the PLI power in the first five wavelet scales was precisely estimated from the spectral distribution of each vector of wavelet coefficients. The values obtained in this way were then used to threshold the wavelet coefficients resulting from the input signal decomposition. For this purpose, the well-known soft thresholding function was used. Hence, those wavelet coefficients below the threshold were set to zero and the remaining ones were reduced by that value. It should be noted that both the input signal and the noise estimated by the notch filtering were decomposed making use of a 2nd-order Coiflet function as mother wavelet (see Figure 1). The denoised EGM was finally reconstructed by applying the IDWT to the thresholded wavelet coefficients.

### 2.3. Performance assessment

For comparison purposes, a notch filter similar to the one used by many commercial recording systems was also replicated through a second-order Butterworth structure. It was designed with a 2 Hz bandwidth centered on the frequency of 50 Hz and applied to the EGM signal in a forward/backward fashion for zero phase distortion.

To quantify both noise reduction and morphology preservation in the resulting EGM signals for the two analyzed algorithms, a signed correlation index (SCI) was computed both in time and frequency domains. This index was used instead of the common Pearson's correlation coefficient, because it takes into consideration the amplitude differences between the two compared signals. From a mathematical point of view, referring to the original clean EGM as $x(n)$ and to the denoised recording as $\widehat{x}(n)$, the SCI was defined as

$$\text{SCI}\left[x(n), \widehat{x}(n)\right] = \frac{1}{N}\sum_{k=1}^{N} x(k) \otimes \widehat{x}(k), \quad (1)$$

where $N$ is the number of samples both for $x(n)$ and $\widehat{x}(n)$ and the operator $\otimes$ is computed as

$$x(n) \otimes \widehat{x}(n) = \begin{cases} 1 & \text{if } |x(n) - \widehat{x}(n)| \leq \xi, \\ -1 & \text{if } |x(n) - \widehat{x}(n)| > \xi. \end{cases} \quad (2)$$

The threshold $\xi$ was experimentally set to 5% of the standard deviation of $x(n)$.

Table 1. Mean and standard deviation values of SCI obtained for the proposed DWT-based denosing algorithm and the reference notch filtering both from time and frequency domains.

| Method | SCI (%) | SIR (dB) | | | | |
|---|---|---|---|---|---|---|
| | | 25 | 20 | 15 | 10 | 5 |
| DWT-based | Time | 94.1 ± 1.6 | 93.6 ± 1.6 | 92.4 ± 1.6 | 89.6 ± 1.7 | 85.9 ± 2.5 |
| | Frequency | 99.5 ± 0.3 | 99.4 ± 0.3 | 99.3 ± 0.4 | 99.1 ± 0.4 | 98.5 ± 0.5 |
| Notch filtering | Time | 79.0 ± 4.1 | 78.9 ± 4.1 | 78.9 ± 4.1 | 78.8 ± 4.0 | 78.3 ± 4.1 |
| | Frequency | 96.4 ± 0.6 | 96.4 ± 0.6 | 96.4 ± 0.6 | 96.3 ± 0.6 | 96.3 ± 0.6 |

## 3. Results

Mean and standard deviation values of SCI obtained for the proposed Wavelet denoising and the reference notch filtering, both from time and frequency domains, are displayed in Table 1. As can be observed, the notch filtering reported a highly stable behavior for every level of SIR. However, it was always worse than the one showed by the proposed DWT-based denoising. In fact, although the values of SCI for the Wavelet denoising reported a slightly decreasing trend with the noise level, they were notably larger than those obtained for the notch fitering, even for a level of SIR of 5 dB. More precisely, compared with the notch filtering, improvements between 4 and 15% were reached by the Wavelet denoising in frequency and time domains, respectively. As a graphical example, Figure 2 displays the resulting signals for both denoising methods, in time and frequency domains, when a SIR of 10 dB was considered.

## 4. Discussion and conclusions

According to some findings reported by previous works [11], the notch filtering has proven to modify substantially the original signal morphology, both in time and frequency domains. Moreover, this method has also been featured by causing a similar alteration in the EGM recording regardless of the noise level, thus completely disturbing even those signals with a very limited presence of PLI. This poor outcome could be easily explained by the massive removal of information around 50 Hz, which is highly relevant in the EGM signal. In fact, rapid deflections originated in the pulmonary veins and Purkinke fibers provoke an EGM bandwidth ranging between 1 and 300 Hz [11].

Contrarily, the proposed Wavelet denoising algorithm has revealed a notably better trade-off between PLI removal and EGM morphology preservation. Thus, even for a very reduced level of SIR of 10 dB, average values of SCI about 90 and 99% were noticed in time and frequency domains, respectively. To this respect, Figure 2 just displays how the EGM is visually preserved after denoising, both in time and frequency domains. As a consequence, the use of this algorithm in routine electrophysiology studies may be helpful for more accurate interpretation of EGM recordings, which could improve current knowledge about the mechanisms of AF.

## Acknowledgements

Research supported by the grants DPI2017-83952-C3 MINECO/AEI/FEDER, UE and SBPLY/17/180501/000411 from Junta de Comunidades de Castilla-La Mancha.

## References

[1] Potter BJ, Le Lorier J. Taking the pulse of atrial fibrillation. The Lancet July 2015;386(9989):113–115.

[2] Zoni-Berisso M, Lercari F, Carazza T, Domenicucci S. Epidemiology of atrial fibrillation: European perspective. Clinical Epidemiology December 2013;6:213–220.

[3] Schnabel RB, Yin X, Gona P, Larson MG, Beiser AS. 50 year trends in atrial fibrillation prevalence, incidence, risk factors, and mortality in the Framingham Heart Study: a cohort study. The Lancet 2015;88(9):815–815.

[4] Schotten U, Dobrev D, Platonov PG, Kottkamp H, Hindricks G. Current controversies in determining the main mechanisms of atrial fibrillation. Journal of Internal Medicine April 2016;279(5):428–438.

[5] Van Wagoner DR, Piccini JP, Albert CM, Anderson ME, Benjamin EJ, Brundel B, Califf RM, Calkins H, Chen PS, Chiamvimonvat N, Darbar D, Eckhardt LL, Ellinor PT, Exner DV, Fogel RI, Gillis AM, Healey J, Hohnloser SH, Kamel H, Lathrop DA, Lip GYH, Mehra R, Narayan SM, Olgin J, Packer D, Peters NS, Roden DM, Ross HM, Sheldon R, Wehrens XHT. Progress toward the prevention and treatment of atrial fibrillation: A summary of the Heart Rhythm Society Research Forum on the Treatment and Prevention of Atrial Fibrillation, Washington, DC, December 9-10, 2013. Heart Rhythm Jan 2015;12(1):e5–e29.

[6] Latchamsetty R, Morady F. Atrial Fibrillation Ablation, volume 69. University of Michigan, Ann Arbor, Ann Arbor, United States, January 2018.

[7] Koutalas E, Rolf S, Dinov B, Richter S, Arya A, Bollmann A, Hindricks G, Sommer P. Contemporary mapping techniques of complex cardiac arrhythmias - Identifying and modifying the arrhythmogenic substrate. Arrhythmia Electrophysiology Review May 2015;4(1):19–27.

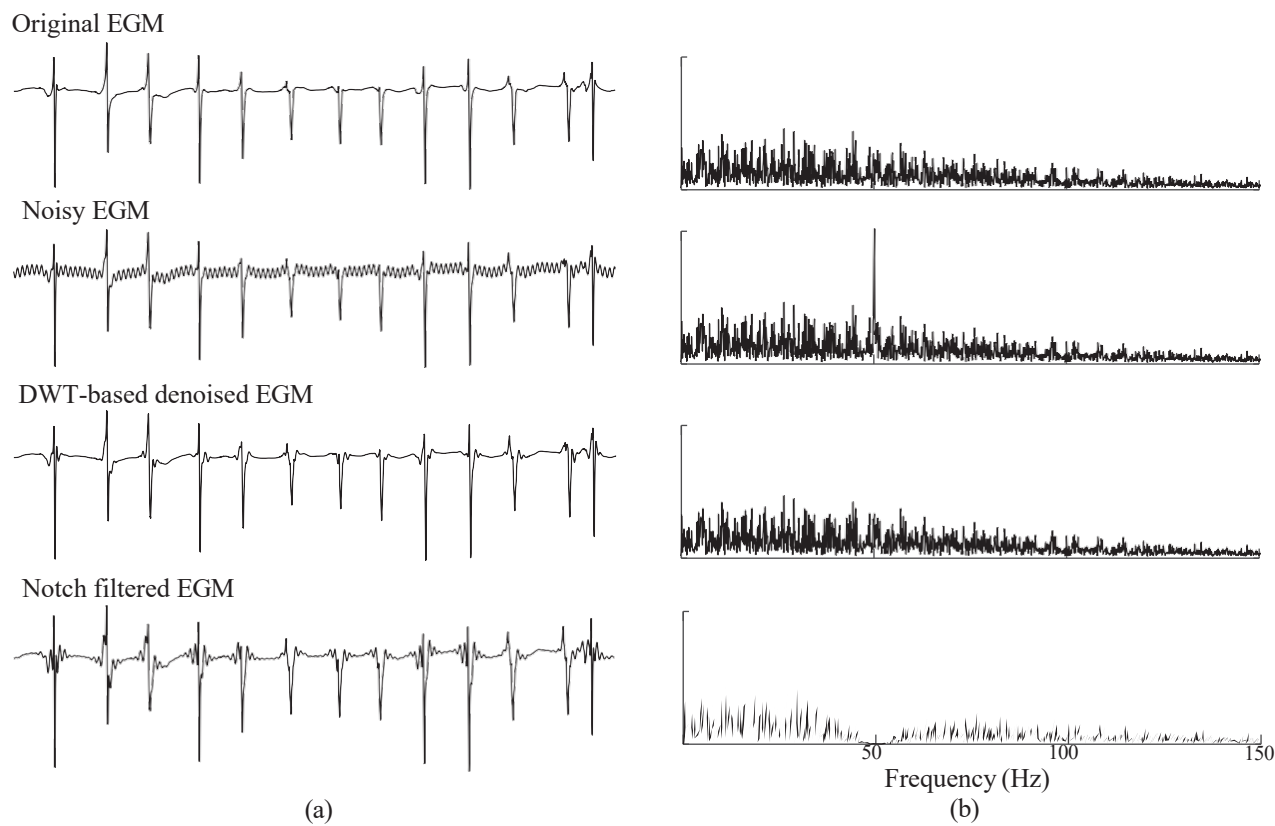

Figure 2. Typical example of the resulting EGM signals for the two analyzed denoising methods both in time (a) and frequency (b) domains, when a SIR of 10 dB was considered.

[8] de Bakker JMT, Wittkampf FHM. The pathophysiologic basis of fractionated and complex electrograms and the impact of recording techniques on their detection and interpretation. Circulation Arrhythmia and Electrophysiology April 2010;3(2):204–213.

[9] Stevenson WG, Soejima K. Recording techniques for clinical electrophysiology. Journal of cardiovascular electrophysiology September 2005;16(9):1017–1022.

[10] Heijman J, Algalarrondo V, Voigt N, Melka J, Wehrens XHT, Dobrev D, Nattel S. The value of basic research insights into atrial fibrillation mechanisms as a guide to therapeutic innovation: A critical analysis. Cardiovascular Research April 2016;109(4):467–479.

[11] Venkatachalam KL, Herbrandson JE, Asirvatham SJ. Signals and signal processing for the electrophysiologist: part I: electrogram acquisition. Circulation Arrhythmia and Electrophysiology November 2011;4(6):965–973.

[12] Venkatachalam KL, Herbrandson JE, Asirvatham SJ. Signals and signal processing for the electrophysiologist: part II: signal processing and artifact. Circulation Arrhythmia and Electrophysiology November 2011;4(6):974–981.

[13] Prutchi D, Norris M. Design and Development of Medical Electronic Instrumentation. A Practical Perspective of the Design, Construction, and Test of Medical Devices. Hoboken, NJ, USA: John Wiley & Sons, Inc., January 2005.

[14] Oesterlein TG, Lenis G, Rudolph DT, Luik A, Verma B, Schmitt C, D ssel O. Removing ventricular far-field signals in intracardiac electrograms during stable atrial tachycardia using the periodic component analysis. Journal of Electrocardiology March 2015;48(2):171–180.

[15] Thamarai P, Adalarasu K. Denoising of EEG, ECG and PPG signals using Wavelet transform. Journal of Pharmaceutical Sciences and Research 2018;10(1):156–161.

[16] Xu L, Yong Y. Wavelet-based removal of sinusoidal interference from a signal. Measurement Science and Technology 2004;15(9):1779.

Address for correspondence:

Miguel Martínez Iniesta
E.I. Industriales, Campus Universitario, 02071, Albacete, Spain
Phone: +34–967–599–200 Ext. 2555
Fax: +34–967–599–224
e–mail: miguel.martinez@uclm.es